\documentclass[preprint]{aastex}

\usepackage{txfonts}
\usepackage{natbib}
\usepackage{graphicx}

\newcommand{\Rnom}{\hbox{$\mathcal{R}^{\mathrm N}_\odot$}}
\newcommand{\Snom}{\hbox{$\mathcal{S}^{\mathrm N}_\odot$}}
\newcommand{\Lnom}{\hbox{$\mathcal{L}^{\mathrm N}_\odot$}}
\newcommand{\Tnom}{\hbox{$\mathcal{T}^{\mathrm N}_\odot$}}
\newcommand{\Mnom}{\hbox{$\mathcal{M}^{\mathrm N}_\odot$}}
\newcommand{\GMnom}{\hbox{$\mathcal{(GM)}^{\mathrm N}_\odot$}}

\newcommand{\REenom}{\hbox{$\mathcal{R}^{\mathrm N}_\mathrm{Ee}$}}
\newcommand{\REpnom}{\hbox{$\mathcal{R}^{\mathrm N}_\mathrm{Ep}$}}
\newcommand{\RJenom}{\hbox{$\mathcal{R}^{\mathrm N}_\mathrm{Je}$}}
\newcommand{\RJpnom}{\hbox{$\mathcal{R}^{\mathrm N}_\mathrm{Jp}$}}
\newcommand{\MEnom}{\hbox{$\mathcal{M}^{\mathrm N}_\mathrm{E}$}}
\newcommand{\MJnom}{\hbox{$\mathcal{M}^{\mathrm N}_\mathrm{J}$}}
\newcommand{\GMEnom}{\hbox{$\mathcal{(GM)}^{\mathrm N}_\mathrm{E}$}}
\newcommand{\GMJnom}{\hbox{$\mathcal{(GM)}^{\mathrm N}_\mathrm{J}$}}

\newcommand{\rev}[1]{{#1}}

\begin{document}

\title{Nominal values for selected solar and planetary quantities: IAU 2015 Resolution B3\altaffilmark{*,\dagger}}

\author{
  Andrej Pr\v{s}a\altaffilmark{1},
  Petr~Harmanec\altaffilmark{2},
  Guillermo Torres\altaffilmark{3},
  Eric Mamajek\altaffilmark{4},
  Martin Asplund\altaffilmark{5},
  Nicole Capitaine\altaffilmark{6},
  J\o rgen Christensen-Dalsgaard\altaffilmark{7},
  \' Eric Depagne\altaffilmark{8,9},
  Margit Haberreiter\altaffilmark{10},
  Saskia Hekker\altaffilmark{11,12},
  James Hilton\altaffilmark{13},
  \rev{Greg Kopp\altaffilmark{14},
  Veselin Kostov\altaffilmark{15},}
  Donald W.~Kurtz\altaffilmark{16},
  Jacques Laskar\altaffilmark{17},
  Brian D.~Mason\altaffilmark{13},
  Eugene F.\ Milone\altaffilmark{18},
  Michele Montgomery\altaffilmark{19},
  Mercedes Richards\altaffilmark{20},
  \rev{Werner Schmutz\altaffilmark{10},}
  Jesper Schou\altaffilmark{21} and
  Susan G.~Stewart\altaffilmark{13}
}
\altaffiltext{*}{Available at {\tt http://www.iau.org/static/resolutions/IAU2015\_English.pdf}.}
\altaffiltext{$\dagger$}{This paper is dedicated to the memory of our dear colleague Dr.~Mercedes Richards, an exquisite interacting binary star scientist and IAU officer who passed away on February 3, 2016.}
\altaffiltext{1}{Villanova University, Dept.~of Astrophysics and Planetary Science, 800 Lancaster Ave, Villanova, PA 19085, USA}
\altaffiltext{2}{Astronomical Institute of the Charles University, Faculty of Mathematics and Physics, V~Hole\v{s}ovi\v{c}k\'ach 2, CZ-180~00~Praha~8, Czech Republic}
\altaffiltext{3}{Harvard-Smithsonian Center for Astrophysics, Cambridge, MA 02138, USA}
\altaffiltext{4}{Department of Physics \& Astronomy, University of Rochester, Rochester, NY, 14627-0171, USA}
\altaffiltext{5}{Research School of Astronomy and Astrophysics, Australian National University, Canberra, ACT 2611, Australia}
\altaffiltext{6}{SYRTE, Observatoire de Paris, PSL Research University, CNRS, Sorbonne Universit\' es, UPMC, LNE, 61 avenue de l’Observatoire, 75014 Paris, France}
\altaffiltext{7}{Stellar Astrophysics Centre, Department of Physics and Astronomy, Aarhus University, Ny Munkegade 120, DK-8000 Aarhus C, Denmark}
\altaffiltext{8}{South African Astronomical Observatory, P.O. Box 9 Observatory, Cape Town, South Africa}
\altaffiltext{9}{Southern African Large Telescope, P.O. Box 9 Observatory, Cape Town, South Africa}
\altaffiltext{10}{Physikalisch-Meteorologisches Observatorium Davos / World Radiation Center, Dorfstrasse 33, Davos, Switzerland}
\altaffiltext{11}{Max-Planck-Institut f\"ur Sonnensystemforschung, Justus-von-Liebig-Weg 3, 37077 G\"ottingen, Germany}
\altaffiltext{12}{Stellar Astrophysics Centre, Department of Physics and Astronomy, Aarhus University, Ny Munkegade 120, 8000 Aarhus C, Denmark}
\altaffiltext{13}{US Naval Observatory, 3450 Massachusetts Ave NW, Washington DC 20392-5420, D.C., USA}
\altaffiltext{14}{\rev{Laboratory for Atmospheric and Space Physics, 1234 Innovation Drive, Boulder, CO  80303-7814, 303-492-6412}}
\altaffiltext{15}{NASA Goddard Space Flight Center, Mail Code 665, Greenbelt, MD, 20771}
\altaffiltext{16}{Jeremiah Horrocks Institute, University of Central Lancashire, Preston PR1 2HE, UK}
\altaffiltext{17}{ASD/IMCCE, CNRS-UMR8028, Observatoire de Paris,  PSL, UPMC, 77 Avenue Denfert-Rochereau, 75014 Paris, France}
\altaffiltext{18}{Prof.\ Emer., Physics \& Astronomy Dept., University of Calgary, 2500 University Dr., N.W., Calgary, AB, T2N~1N4, Canada}
\altaffiltext{19}{University of Central Florida, Physics Department, 4000 Central Florida Blvd., Orlando, FL, 32816, USA}
\altaffiltext{20}{Department of Astronomy \& Astrophysics, Pennsylvania State University, University Park, PA 16802, USA}
\altaffiltext{21}{Max Planck Institute for Solar System Research, Justus-von-Liebig-Weg 3, 37077 G\" ottingen, Germany}

\shortauthors{Pr\v{s}a et al.}
\shorttitle{Nominal solar and planetary quantities}

\date{Received April 1, 2016; accepted mmmm dd, yyyy}

\begin{abstract}
In this brief communication we provide the rationale \rev{for,} and the outcome of the International Astronomical Union (IAU) resolution vote at the XXIX$^{\rm th}$ General Assembly in Honolulu, Hawaii, in 2015, on recommended nominal conversion constants for selected solar and planetary properties. The problem addressed by the resolution is a lack of established conversion constants between solar and planetary values and SI units: \rev{a missing standard has caused a proliferation of solar values (e.g.,}~solar radius, solar irradiance, solar luminosity, solar effective temperature and solar mass parameter) \rev{in the literature, with cited} solar values typically based on best estimates at the time \rev{of paper writing. As} precision of observations \rev{increases, a set} of consistent values becomes increasingly important. \rev{To address this, an} IAU Working Group on Nominal Units for Stellar and Planetary Astronomy \rev{formed in 2011, uniting} experts from the solar, stellar, planetary, exoplanetary and fundamental astronomy, as well as from general standards fields to converge on optimal values for nominal conversion constants. The \rev{effort resulted in the} IAU 2015 Resolution B3, passed at the IAU General Assembly by a large majority. \rev{The resolution} recommends the use of \emph{nominal} solar and planetary values, which are by definition \emph{exact} and \rev{are expressed in SI units}. These nominal values should be understood as conversion factors only, not as the true solar/planetary properties \rev{or current best estimates}. Authors and journal editors are urged to join in using the standard values set forth by this resolution in future work and publications to help minimize \rev{further} confusion.
\end{abstract}


\section{Introduction}

It is customary in stellar astrophysics to express properties of stars in terms of solar values, for example $2.2 \, M_\odot$, $1.3 \, R_\odot$, etc. The problem arises when these quantities need to be transformed to the International System of units (SI). More often than not, authors do not report the conversion constants used in their work, and the differences that stem from using different values are in some instances no longer negligible. \citet{harmanec2011} raised this issue and demonstrated its impact on several formulae widely used in binary star astrophysics. Analogously, planetary and exoplanetary scientists commonly express planetary properties in terms of Earth or Jupiter values. This custom is plagued by the same problem. As a simple demonstration, providing a planet size of $0.7538 \pm 0.0025 \,R_\mathrm{J}$ (as happens to be the case for Kepler 16; \citealt{doyle2011}) can be interpreted in (at least) three ways, depending on what $R_\mathrm J$ is assumed to be: mean radius, equatorial radius, or polar radius. According to \citet{archinal2011}, the mean (m), equatorial (e) and polar (p) radii of Jupiter correspond to the layer at $1\,$bar of pressure and are, respectively, $R_\mathrm{Jm} = 69\,911 \pm 6$\,km, $R_\mathrm{Je} = 71\,492 \pm 4$\,km, and $R_\mathrm{Jp} = 66\,854 \pm 10$\,km. Thus, the size of Kepler 16 could be interpreted as either of $52\,699 \pm 175$\,km, $53\,891 \pm 179$\,km, or $50\,395 \pm 167$\,km. Clearly, the systematic error due to an unspecified conversion constant dominates the uncertainty budget: $\sim$6.5\% compared to the model uncertainty of 0.3\%, which represents a factor of more than 20.

\section{Steps toward the Resolution}

A viable solution to the problem of notably different values for the solar, terrestrial, and jovian properties used in the literature and in software is to \rev{abandon} the use of \emph{measured} values and introduce \rev{instead} the use of \emph{nominal} conversion constants. These conversion constants should be chosen to be close to the current measured values (current best estimates) for convenience, since it seems unlikely that the community would be eager to adopt a significantly different set of measurement scales, which would imply the loss of backwards consistency and the loss of familiar relations between, for example, effective temperature, mass, radius, and luminosity. Although the constants are chosen to be close to measured values by design, they should not be confused for actual solar/planetary properties (current best estimates); they are simply conversion constants between a convenient measure for stellar/planetary-size bodies, and the same properties in SI units. The nominal units are designed to be useful ``rulers'' for the foreseeable future. A classical example is the standard acceleration due to Earth's gravity, $g_n$, set to $g_n = 9.80665$\,m\,s${}^{-2}$ (originally in cgs units) by the 3rd General Conference on Weights and Measures (Conf\' erence g\' en\' erale des poids et mesures; CGPM) in 1901. Whereas the true value of $g$ varies over the surface of the Earth, $g_n$ is its internationally recognized nominal counterpart, which has remained unchanged for over a century.

In 2011, when the first formal proposal was presented by \citeauthor{harmanec2011}, only the solar radius and solar luminosity were nominalized. The reason for this was that other parameters, such as the mass, can be measured to a much higher precision as coupled quantities, i.e., the solar mass parameter $GM_\odot$. Because the uncertainty in $G$ is \rev{five} orders of magnitude larger than the uncertainty in $GM_\odot$ \citep[e.g.,][]{petit2010, luzum2011}, the conversion from $M/M_\odot$ to SI would suffer from the uncertainty in $G$. To make that abundantly clear, we proposed the use of the $2010$ superscript to denote the CODATA year of the values used for the physical constants. However, as these quantities carried the propagated uncertainty with them, this was not a sufficiently robust solution. The proposal met with general approval, most notably by a near-unanimous participant vote at the IAU Symposium 282 in Tatranska Lomnica, Slovakia \citep{prsa2012, richards2012}, and at the Division~A business meeting at the IAU General Assembly in Beijing, China, but a clear case for further refinement was quickly established.

In the following three years, the IAU instituted a Working Group on Nominal Units for Stellar and Planetary Astronomy (hereafter WG) under the auspices of Divisions G and A and with support from Divisions F, H, and J. The WG was chaired by Dr.\ P.\ Harmanec until 2015 and by Dr.\ E.\ Mamajek since 2015, and co-chaired by Drs.\ A.\ Pr\v sa and G.\ Torres. The WG \rev{brought together} 23 experts from around the world and from different fields to discuss and further refine the proposal, with the goal of writing the Resolution draft and putting it to the member-wide vote at the 2015 IAU General Assembly in Honolulu, Hawaii. During the same \rev{period} the WG also addressed the standardization of the absolute and apparent bolometric magnitude scales, which resulted in an independent resolution proposal B2 (Mamajek et al.~2016, in preparation) that passed the vote at the same time as the resolution on nominal conversion constants. Both resolutions passed the XXIX${}^\mathrm{th}$ IAU General Assembly vote by a large majority.

In the next two sections we provide the Resolution and the rationale for the proposed values of the nominal conversion constants. In Appendix~\ref{app:eqs} we provide an update of the list of formulae from \citet{harmanec2011}, with the current nominal values. \emph{The numerical values proposed by \citet{harmanec2011} are superseded by the present paper and should no longer be used}.

\section{IAU 2015 Resolution B3}

In this section we reproduce the recommendations of the Resolution essentially verbatim\footnote{Also available at {\tt http://www.iau.org/static/resolutions/IAU2015\_English.pdf} and \\ {\tt http://arXiv.org/abs/1510.07674}.}, with minimal typesetting adaptations.

Noting that (1) neither the solar nor the planetary masses and radii are secularly constant and that their instantaneous values are gradually being determined more precisely through improved observational techniques and methods of data analysis, and (2) that the common practice of expressing the stellar and planetary properties in units of the properties of the Sun, the Earth, or Jupiter inevitably leads to unnecessary systematic differences that are becoming apparent with the rapidly increasing accuracy of spectroscopic, photometric, and interferometric observations of stars and extrasolar planets, and (3) that the universal constant of gravitation $G$ is currently one of the least precisely determined constants, whereas the error in the product $GM_\odot$ is five orders of magnitude smaller \citep[and references therein]{petit2010}, the Resolution makes the following recommendations applicable to all scientific publications in which accurate values of basic stellar or planetary properties are derived or quoted:
\begin{enumerate}
\item that whenever expressing stellar properties in units of the solar radius, total solar irradiance, solar luminosity, solar effective temperature, or solar mass parameter, the nominal values $\Rnom$, $\Snom$, $\Lnom$, $\Tnom$, and $\GMnom$ be used, respectively, which are by definition exact and are expressed in SI units. These \emph{nominal} values should be understood as conversion factors only --- chosen to be close to the current commonly accepted estimates (see Table \ref{tab:nompars}) --- not as the true solar properties. Their consistent use in all relevant formulae and/or model calculations will guarantee a uniform conversion to SI units. Symbols such as $L_\odot$ and $R_\odot$, for example, should only be used to refer to actual estimates of the solar luminosity and solar radius (with uncertainties);

\item that the same be done for expressing planetary properties in units of the equatorial and polar radii of the Earth and Jupiter (i.e., adopting nominal values $\REenom$, $\REpnom$, $\RJenom$, and $\RJpnom$, expressed in meters), and the nominal terrestrial and jovian mass parameters $\GMEnom$ and $\GMJnom$, respectively (expressed in units of m${}^3$ s${}^{-2}$). Symbols such as $GM_\mathrm E$, listed in the IAU 2009 system of astronomical constants \citep{luzum2011}, should be used only to refer to actual estimates (with uncertainties);

\item that the IAU (2015) System of Nominal Solar and Planetary Conversion Constants be adopted as listed in Table \ref{tab:nompars},

\item that an object's mass be quoted in nominal solar masses $\Mnom$ by taking the ratio $(GM)_\mathrm{object}/\GMnom$, or in corresponding nominal terrestrial and jovian masses, $\MEnom$ and $\MJnom$, respectively, dividing by $\GMEnom$ and $\GMJnom$;

\item that if SI masses are explicitly needed, they should be expressed in terms of $(GM)_\mathrm{object}/G$, where the estimate of the Newtonian constant $G$ should be specified in the publication (for example, the 2014 CODATA value is $G = (6.67408 \pm 0.00031) \times 10^{11}$\,m${}^3$ kg${}^{-1}$ s${}^{-2}$); and

\item that if nominal volumes are needed, nominal terrestrial volumes be derived as $4\pi \REenom^2 \REpnom / 3$, and nominal jovian volumes as $4\pi \RJenom^2 \RJpnom / 3$.
\end{enumerate}

\begin{table}[t]
\begin{center}
\begin{tabular}{lclclcl}
\hline \hline
\multicolumn{3}{c}{SOLAR CONVERSION CONSTANTS}                         & \phantom{space} & \multicolumn{3}{c}{PLANETARY CONVERSION CONSTANTS} \\
\hline
$1 \Rnom$  & $=$ & $6.957 \times 10^8$\,m                              & \phantom{space} & $1 \REenom$ & $=$ & $6.3781 \times 10^6$\,m \\
$1 \Snom$  & $=$ & $1361$\,W\,m${}^{-2}$                               & \phantom{space} & $1 \REpnom$ & $=$ & $6.3568 \times 10^6$\,m \\
$1 \Lnom$  & $=$ & $3.828 \times 10^{26}$\,W                           & \phantom{space} & $1 \RJenom$ & $=$ & $7.1492 \times 10^7$\,m \\
$1 \Tnom$  & $=$ & $5772$\, K                                          & \phantom{space} & $1 \RJpnom$ & $=$ & $6.6854 \times 10^7$\,m \\
$1 \GMnom$ & $=$ & $1.327\,124\,4 \times 10^{20}$\,m${}^3$\,s${}^{-2}$ & \phantom{space} & $1 \GMEnom$ & $=$ & $3.986\,004 \times 10^{14}$\,m${}^3$\,s${}^{-2}$ \\
           &     &                                                     & \phantom{space} & $1 \GMJnom$ & $=$ & $1.266\,865\,3 \times 10^{17}$\,m${}^3$\,s${}^{-2}$ \\
\hline
\end{tabular}
\caption{
\label{tab:nompars}
Nominal solar and planetary conversion constants set forth by IAU 2015 Resolution B3. Although chosen to be as close to the measured quantities as feasible given the observational uncertainties for practical reasons, these values should \emph{not} be considered the true solar/planetary properties. They should be understood as conversion values only.
}
\end{center}
\end{table}

\section{The Rationale and Considerations}

As nominal conversion constants represent a new set of units of measure rather than any measured quantity itself, the values need not match their physical counterparts, though as stated earlier, it is convenient that they do so. Thus, substantial deliberation went into the selection of the proposed values, and various considerations and the rationale behind these values are presented here:

\begin{description}
\item[The value of the nominal solar radius $\Rnom$.] The chosen value corresponds to the solar photospheric radius suggested by \citet{haberreiter2008}, defined to be where the Rosseland optical depth $\tau = 2/3$. This study resolved the long-standing discrepancy between the seismic and photospheric solar radii. The nominal value ($6.957 \times 10^8$\,m) is the rounded \citeauthor{haberreiter2008}~value ($695\,658 \pm 140$\,km) within the uncertainty. This $\Rnom$ value is very close to the value adopted by \citet{torres2010} in their compilation of updated radii of well observed eclipsing binary systems, and differs slightly from the nominal solar radius tentatively proposed by \citet{harmanec2011} and \citet{prsa2012}.

\item[The value of the nominal total solar irradiance $\Snom$.] The chosen value corresponds to the mean total electromagnetic energy from the Sun, integrated over all wavelengths, incident per unit area and per unit time at a distance of 1\,au. The Total Solar Irradiance (TSI; \citealt{willson1978}) is variable at the $\sim$0.08\% ($\sim$1\,W\,m${}^{-2}$) level and may be variable at slightly larger amplitudes over timescales of centuries. Modern spaceborne TSI instruments are calibrated absolutely to SI irradiance standards at the 0.03\% level \citep{kopp2014}. The TIM/SORCE experiment established a lower TSI value than previously reported based on the fully characterized TIM instrument \citep{kopp2005, kopp2011}. This revised TSI scale was later confirmed by PREMOS/PICARD, the first spaceborne TSI radiometer that was irradiance-calibrated in vacuum at the TSI Radiometer Facility (TRF) with SI-traceability prior to launch \citep{schmutz2013}. The ACRIM3/ACRIMSat \citep{willson2014}, VIRGO/SoHO \citep{frohlich1997}, and TCTE/STP-Sat3\,\footnote{{\tt http://lasp.colorado.edu/home/tcte/}} flight instruments are now consistent with this new TSI scale within instrument uncertainties, with the DIARAD, ACRIM3, and VIRGO having made post-launch corrections and the TCTE having been validated on the TRF prior to its 2013 launch. \rev{Using any of the available TSI composites,} the Cycle 23 observations with these experiments are consistent with a mean TSI value of \rev{$S_\odot = 1361 \pm 0.5$\,W\,m${}^{-2}$. This uncertainty reflects absolute accuracies of the latest TSI instruments as well as uncertainties in assessing a secular trend in TSI over solar cycle 23 using older measurements and is fully consistent with the uncertainty reported by \citet{kopp2011}}. Our adopted value of $\Snom$ corresponds to this solar cycle 23-averaged TSI.

\item[The value of the nominal solar luminosity $\Lnom$.] The chosen value corresponds to the mean solar radiative luminosity. The best estimate of the mean solar luminosity $L_\odot$ was calculated using the solar cycle-averaged TSI (see above) and the IAU 2012 definition of the astronomical unit. Resolution B2 of the XXVIII General Assembly of the IAU in 2012 defined the astronomical unit to be a nominal unit of length equal to 149\,597\,870\,700\,m. Using the current best estimate of the TSI, we arrive at the current best estimate of the Sun's mean radiative luminosity of $L_\odot = 4\pi (1~\mathrm{au})^2 S_\odot = (3.8275 \pm 0.0014) \times 10^{26}$\,W. The Resolution adopts a rounded value of this current best estimate.

\item[The value of the nominal solar effective temperature $\Tnom$.] The current best estimate for the solar effective temperature is derived from the Stefan-Boltzmann law, using the current best estimates for the solar photospheric radius and solar radiative luminosity, and the CODATA 2014 value for the Stefan-Boltzmann constant $\sigma = (5.670367 \pm 0.000013) \times 10^{-8}$\,W\,m${}^{-2}$\,K${}^{-4}$, yielding $T_\odot = 5772.0 \pm 0.8$\,K. The chosen value for $\Tnom$ is a truncated value of $T_\odot$, consistent within the uncertainty.

\item[The value of the nominal solar mass parameter $\GMnom$.] In solar and planetary astronomy, time is typically referenced in one of two coordinate time scales \citep{klioner2006}: Barycentric Coordinate Time (Temps coordonn\' ee barycentrique; TCB) and Barycentric Dynamical Time (Temps dynamique barycentrique; TDB; defined by IAU 2006 Resolution B3). TDB includes relativistic corrections due to time dilation and it can be written as a linear transformation of TCB. The nominal value of $\GMnom$ is based on the best available measurement \citep{petit2010} but rounded to the precision to which both TCB and TDB values agree \citep{luzum2011}. This precision is considered to be sufficient for most applications in stellar and exoplanetary research for the foreseeable future.

\item[The values of the nominal terrestrial radii $\REenom$ and $\REpnom$.] These parameters correspond to the Earth's ``zero-tide'' equatorial and polar radii, respectively, adopted from the 2003 and 2010 International Earth rotation and Reference system Service (IERS) Conventions \citep{mccarthy2004, petit2010}, the IAU 2009 system of astronomical constants \citep{luzum2011}, and the IAU Working Group on Cartographic Coordinates and Rotational Elements \citep{archinal2011}. If the terrestrial radius is not explicitly qualified as equatorial or polar, it should be understood that nominal terrestrial radius refers specifically to the equatorial radius, following common usage in the literature.

\item[The values of the nominal jovian radii $\RJenom$ and $\RJpnom$.] These parameters correspond to the one-bar equatorial and polar radii of Jupiter, respectively, adopted by the IAU Working Group on Cartographic Coordinates and Rotational Elements 2009 \citep{archinal2011}. If the jovian radius is not explicitly qualified as equatorial or polar, it should be understood that nominal jovian radius refers specifically to the equatorial radius, following common usage in the literature.

\item[The values of the nominal mass parameters $\GMEnom$ and $\GMJnom$.] The \emph{nominal terrestrial mass parameter} is ad\-op\-t\-ed from the geocentric gravitational constant from the IAU 2009 system of astronomical constants \citep{luzum2011}, but rounded to the precision within which its TCB and TDB values agree (cf., the discussion for $\GMnom$ above). The \emph{nominal jovian mass parameter} is calculated based on the mass parameter for the Jupiter system from the IAU 2009 system of astronomical constants \citep{luzum2011}, subtracting the contribution from the Galilean satellites \citep{jacobson2000}. The quoted value is rounded to the precision within which the TCB and TDB values agree, and the uncertainties in the masses of the satellites are negligible in some instances.
\end{description}

\section{Discussion and Conclusions}

The Resolution is part of an ongoing effort to introduce a consistent and robust set of ``rulers'' to be used in modern stellar and planetary astrophysics. While the overwhelming vote of confidence and the passed Resolution help a great deal, the next step is for the community to adopt the practice of using these values in all related work and publications. Examples of important uses of the nominal conversion constants, particularly those pertaining to the Sun, include the calibration of stellar evolution models and the tabulation of evolutionary tracks and isochrones derived from those models. In particular, the mass $M_\mathrm{model}/\Mnom$ in nominal units assigned to evolutionary tracks can be obtained as $(G M_{\rm model})/\GMnom$, where $G$ (to be explicitly specified in the publication; see, e.g., CODATA 2014 constants; \citealt{mohr2015}) and the mass $M_{\rm model}$ of the model are in SI units. The grid of stellar models of Choi et al.\ (2016, in press), based on the MESA code \citep{paxton2011, paxton2013, paxton2015}, has already adopted the recommended values, and other groups are encouraged to do the same. Future libraries of synthetic spectra should also be based on a solar calibration using the recommended solar conversion constants. The eclipsing binary modeling code PHOEBE (\citealt{prsa2005}, Pr\v sa et al.~2016, in preparation) has also been updated to use the recommended values. Developers of other codes for binary star orbital solutions (photometric, spectroscopic, astrometric, etc.) are encouraged to adopt the new conversion constants as well, so that future stellar mass and radius measurements for binary stars are reported on a homogeneous system.

The conversion constants established by this Resolution are a subset of nominal conversion constants defined in other IAU resolutions that the community is encouraged to use. For most stellar and planetary purposes, the distance scale is given in astronomical units or parsecs. In 2012, the IAU passed a resolution that defines the astronomical unit as 149\,597\,870\,700\,m. Along with it, IAU 2015 Resolution B2 adopted the definition of the parsec to be exactly $648000/\pi$\,au (\citealt{binney2008}; see also \citealt{cox2000}). Since $\pi$ is irrational, the length of one parsec cannot be rational, but it still is an \emph{exact} number, $3.085\,677\,581\,491\,\cdots\, \times 10^{16}$\,m.

Other challenges still remain unresolved. Of notable importance are the definitions of the semi-major axis and of the orbital period of binary, multiple and exoplanetary systems. General relativity causes corrections of order $(v/c)^2$, which for the Earth are about 1 part in $10^8$. There is \rev{as of yet} no clear consensus on such spatial or temporal references. \rev{This leads to questions such as whether} the semi-major axis \rev{should} be \rev{reported} in barycentric coordinates, photocentric coordinates, or Jacobi coordinates, \rev{whether} the orbital period \rev{should} be measured as sidereal, synodic, or with respect to periapsis passage, how \rev{all of this is influenced by perturbations from} other orbital bodies, etc. Another notable challenge is that of bolometric corrections. The definition of the zero-point of the bolometric magnitude scale has been set ($M_\mathrm{bol} = 0$ corresponds exactly to $L = 3.0128 \times 10^{28}$\,W; cf., IAU 2015 Resolution B2; Mamajek et al.~2016, in preparation), but bolometric corrections still need to be nominalized \citep{torres2010b}. Dedicated experts need to address these issues and propose a draft resolution in the near future.

As noted earlier, the nominal conversion constants were chosen to be close to the corresponding current best estimates. In consequence, the nominal solar effective temperature, nominal solar luminosity, and nominal solar radius are mutually consistent when using the current best estimate of the Stefan-Boltzmann constant $\sigma$. However, nominal units do not \emph{need} to be consistent with any physical laws --- they do not violate them because they are merely a set of ``rulers'', and are close to the current best estimates for convenience only. Whereas the current best estimates will change in the future, the nominal values need not.

In parallel, the International Committee for Weights and Measures (Comit\' e international des poids et mesures; CIPM) has proposed a revised formal definition of the SI base units, which are currently under revision and will likely be adopted at the 26th General Conference on Weights and Measures (Conf\' erence g\' en\' erale des poids et mesures; CGPM) in the Fall of 2018. The basis of the proposal is the redefinition of the kilogram, ampere, kelvin, and mole by choosing exact numerical values for the Planck constant, the elementary electric charge, the Boltzmann constant, and the Avogadro constant, respectively. The meter and candela are already defined by physical constants and it is only necessary to edit their present definitions.

In this paper we have provided the motivation, the \rev{history}, and the rationale for IAU 2015 Resolution B3 on nominal solar and planetary conversion constants that was passed at the XXIX$^{\rm th}$ IAU General Assembly in Honolulu, Hawaii, in 2015. We encourage authors --- as well as journal editors --- to join us in using the standard values set forth by the Resolution in future work and publications to help minimize confusion.

\bigskip

\noindent
A technical note on the use of nominal conversion constants in the typesetting language \LaTeX{}: to obtain the symbols used in this communication, add the following \LaTeX~macros in the preamble, replacing `Q' with the appropriate symbol:
\begin{verbatim}
\newcommand{\Qnom}  {\hbox{$\mathcal{Q}^{\rm N}_\odot$}}
\newcommand{\QEenom}{\hbox{$\mathcal{Q}^{\rm N}_\mathrm{Ee}$}}
\newcommand{\QEpnom}{\hbox{$\mathcal{Q}^{\rm N}_\mathrm{Ep}$}}
\newcommand{\QJenom}{\hbox{$\mathcal{Q}^{\rm N}_\mathrm{Je}$}}
\newcommand{\QJpnom}{\hbox{$\mathcal{Q}^{\rm N}_\mathrm{Jp}$}}
\end{verbatim}
In IDL, the symbols can be obtained using the following markup:
\begin{verbatim}
Rnom='!13R!S!D!9n!R!N!U!6N !N'
Snom='!13S!S!D!9n!R!N!U!6N !N'
Lnom='!13L!S!D!9n!R!N!U!6N !N'
Tnom='!13T!S!D!9n!R!N!U!6N !N'
Mnom='!13M!S!D!9n!R!N!U!6N !N'
GMnom='!13(GM)!S!D!9n!R!N!U!6N !N'
\end{verbatim}
It is our hope that symbols for nominal conversion constants will be provided by journal style files in the future.

\begin{acknowledgements}
We kindly acknowledge discussions with Philip Bennett, Wolfgang Finsterle, William Folkner \rev{and} Hugh Hudson. We further acknowledge remarkable work by Drs.~Allen, Cox and collaborators on ''Astrophysical Quantities``. A.\ Pr\v sa acknowledges support by Villanova University's Summer Fellowship grant. The research of P.~Harmanec was supported by the grants P209/10/0715 and GA15-02112S of the Czech Science Foundation. G.\ Torres acknowledges partial support from NSF award AST-1509375. E.\ Mamajek acknowledges support from NSF AST award 1313029, and NASA's NExSS program. J.\ Christensen-Dalsgaard acknowledges funding for the Stellar Astrophysics Centre that is provided by The Danish National Research Foundation (Grant DNRF106). S.\ Hekker acknowledges funding from the European Research Council under the European Community's Seventh Framework Programme (FP7/2007-2013) / ERC grant agreement no 338251 (StellarAges). The research leading to these results has received funding from the European Community’s Seventh Framework Programme (FP7/2007-2013) under Grant Agreement N${}^\mathrm{o}$ 313188 (SOLID, http://projects.pmodwrc.ch/solid/).
\end{acknowledgements}

\bibliographystyle{aa}
\bibliography{units}

\appendix

\section{Practical guide to using nominal conversion constants} \label{app:eqs}

We present here some practical examples of how to correctly apply the nominal units. Table~\ref{units} contains a list of units, variables, and notation used, and Table~\ref{formu} lists various frequently used formulae for single stars and two-body systems and the numerical values of the constants involved, in nominal units. These values supersede those tentatively suggested by Harmanec \& Pr\v{s}a (2011). {\sl We encourage researchers to incorporate all relevant physical constants and nominal values defined in IAU 2015 resolutions B2 and B3 into their computer programs, and to employ conversion formulae such as those given in the tables to carry out their calculations. This will ensure consistency with the nominal units recommended by the IAU 2015 B2 and B3 resolutions to the level of the computer's numerical accuracy.} The values in the 4$^\mathrm{th}$ column of Table~\ref{formu} are provided as a numerical illustration of conversion expressions provided in column 3 only; we do \emph{not} recommend using them in software implementations or publications. Only the nominal values should be implemented and used explicitly.

The use of nominal units is strongly preferred for the analysis of observational data, such as when solving light curves and radial velocity curves \rev{of binary and multiple systems}. We note that in all of the relevant formulae the mass never appears separately but always in combination (as a product) with the gravitational constant, i.e., as one component of the mass parameter $\mathcal{GM}$. Thus, the SI (or cgs) unit of mass is irrelevant, allowing stellar and exoplanetary masses to be expressed in terms of nominal solar (\Mnom), jovian (\MJnom), or terrestrial (\MEnom) units {\sl without the need for the exact conversion factor to SI or cgs units.} We illustrate this usage with the example of Kepler's third law in a two-body system. The expression for the semi-major axis \rev{$a$} expressed in nominal solar units for radius (\Rnom), with masses $M_1$ and $M_2$ expressed in nominal solar units and the period $P$ in days, and the conversion factor from units of seconds to units of days, $86400$~s~day$^{-1}$, can be written as:
\begin{eqnarray}
a^3 [\Rnom^3] & = &\left({\GMnom\over{4\pi^2(\Rnom)^3}}\right)(86400\,{\rm s}\,{\rm day}^{-1})^2P^2(M_1+M_2)\nonumber\\
& = &\left({{1.3271244 \times 10^{20}\,{\rm m}^3\,{\rm s}^{-2}}\over{4\pi^2(6.957 \times 10^8\,{\rm m})^3}}\right) (86400\,{\rm s}\,{\rm day}^{-1})^2 P^2(M_1+M_2) \nonumber\\
& = &  (74.52695 \cdots) P^2 (M_1 + M_2).
\end{eqnarray}

For the many-body case, instantaneous Keplerian elements can be determined with respect to a specified origin. The choice of center is likely to be application-dependent, and may even change within a given system. With radial-velocity curves, for example, the center of mass would be the preferred origin. If a three-body system has two bodies in close association and a third in a wider orbit, the radial-velocity curves of the close pair would be best served by using the center of mass of that pair alone, while the radial-velocity curve of the third member would be given with respect to the center of mass of the entire system. In all cases it is critical that authors fully describe the
reference frames they are using in order to avoid confusion.

The situation is less clear for stellar interior models. The equations of the stellar interior structure require explicit values for the mass and the constant of gravitation, $G$, in SI (or cgs) units. The IAU Working Group on Numerical Standards for Fundamental Astronomy, NSFA, \citep{luzum2011} recommends:
\begin{eqnarray}
G = 6.67428 \times 10^{-11}\ {\rm m}^3{\rm kg}^{-1}{\rm s}^{-2}.\nonumber
\end{eqnarray}
This is the CODATA 2006 value for $G$ \citep{mohr2008}, and yields a nominal solar mass of:
\begin{eqnarray}
\Mnom = 1.988416 \times 10^{30}\ {\rm kg}. \nonumber
\end{eqnarray}
The slightly different CODATA 2014 value of the gravitational constant \citep{mohr2015} is:
\begin{eqnarray}
G = (6.67408 \pm 0.00031) \times 10^{-11}\ {\rm m}^3{\rm kg}^{-1}{\rm s}^{-2},\nonumber
\end{eqnarray}
which yields a solar mass of:
\begin{eqnarray}
M_\odot & = & (1.988475 \pm 0.000092) \times 10^{30}\ {\rm kg}.\nonumber
\end{eqnarray}
There is currently a large uncertainty in the value $G$, with different researchers arriving at values that differ by several times the formal errors. The NSFA has decided that, for the sake of consistency, it would continue recommending the older value. {\sl Therefore, again it is very important for those calculating stellar interior models to explicitly state the SI values adopted for the gravitational constant $G$ and the model solar mass $M_\odot$, taking care that their chosen values satisfy} $GM_\odot$ = \GMnom.

\bigskip
\noindent Technical remarks:

\begin{enumerate}
\item The numerical constants in the tables are printed to high enough precision to minimize round-off and machine errors in routine calculations. Users working at much higher precision will need to account for other physical effects including those requiring relativistic corrections.  For example, the effect of relativity on the apparent position of a body viewed from the Earth is a few parts in $10^8$ \citep[see][Section B, Reduction of Celestial Coordinates]{almanac}.

\item Resolution B2 of the 2012 IAU General Assembly adopted an exact value for the astronomical unit, au. The notes to Resolution B2 of the 2015 IAU General Assembly define the parsec to also be an exact value, 1~pc = $648000\ \pi^{-1}$~au.

\item The constant numerical value in the formula for stellar bolometric magnitude\rev{s} complies with the values given in the 2015 IAU Resolution B2. A radiation source with absolute bolometric magnitude $M_{\rm Bol}=0$~mag is assumed to have a radiative power of exactly $L_0=3.0128 \times 10^{28}$~W so that the bolometric magnitude $M_{\rm Bol}$ for a source of luminosity $L$, in watts, is:
\begin{eqnarray}
M_{\rm Bol} & = & -2.5\log(L/L_0)\nonumber\\
 & = & -2.5\log L + 71.197425 \cdots
\end{eqnarray}
The nominal solar luminosity is $\Lnom = 3.828 \times 10^{26}$~W, which, given the adopted IAU bolometric magnitude zero point, corresponds approximately to $M_{\rm Bol\odot} = 4.74$~mag. The error in the conversion constant arises from the current uncertainty in the Stefan-Boltzmann constant $\sigma$.

\item Table~\ref{formu} includes the formula to derive the mass function for single-lined spectroscopic binaries (either stellar binaries or a star with an exoplanet), defined as:
\begin{equation}
f_{\rm j}(M_1,M_2) ={4\pi^2\over{G}}{a_{\rm j}^3\sin^3i\over{P^2}} =
{M_{\rm 3-j}^3\sin^3i\over{(M_1+M_2)^2}}
\end{equation}
for $j = 1$ or $2$, based on the spectroscopic elements $P$, $K_{\rm j}$, and $e$. This can be re-arranged to obtain the expression for the mass of the object that is invisible in the spectra, usually component 2, commonly expressed as:
\begin{eqnarray}
M_2\sin i & = & 0.00469686\cdots\ K_1 P^{1/3} (M_1+M_2)^{2/3}\sqrt{(1-e^2)}.
\end{eqnarray}
With an estimate of the mass $M_1$ and the inclination angle $i$ it is then possible to solve the above equation iteratively, beginning with a trial value of $M_2 < M_1$. This approach is also convenient for extrasolar planets since it is possible to start the iteration with $M_2=0$.
\end{enumerate}

\begin{table*}
\begin{center}
\caption{Summary of units and quantities used as variables
in various formul\rev{ae}\ listed in Table~\ref{formu}.}
\label{units}
\begin{tabular}{l c c}
\noalign{\smallskip}\hline\hline\noalign{\smallskip}
\multicolumn{1}{c}{Quantity} & Symbol & Unit\\
\noalign{\smallskip}\hline\noalign{\smallskip}
stellar mass & $M$ & \Mnom\\
stellar mass parameter & $GM$ & \GMnom\\
radius & $R$ & \Rnom\\
luminosity & $L$ & \Lnom\\
luminosity for $M_{\rm Bol}=0$ mag &$L_0$&W\\
astronomical unit& $au$ & m\\
parsec&$pc$&m\\
stellar parallax &$p$& arcsec\\
stellar angular diameter&$\theta$& arcsec\\
orbital period & $P$ & days\\
rotational period& $P_\mathrm{rot}$& days\\
orbital eccentricity & $e$ & -- \\
inclination of orbit or axis of rotation& $i$ & degrees or radians \\
distance of the orbiting component from the center of mass&$a_{1,2}$&\Rnom\\
semi-major axis &$a = a_1+a_2$& \Rnom\\
equatorial rotational velocity& $V_{\rm eq}$ &km\,s$^{-1}$ \\
Keplerian (break-up) velocity&$V_{\rm Kepler}$&km\,s$^{-1}$ \\
effective temperature&$T_\mathrm{eff}$&K\\
surface gravity&$g$&cm\,s$^2$\\
\rev{absolute} bolometric magnitude&$M_{\rm Bol}$&mag\\
\noalign{\smallskip}\hline
\end{tabular}
\end{center}
\end{table*}

\begin{deluxetable}{lrll}
\rotate
\tabletypesize{\scriptsize}
\tablecolumns{4}
\tablewidth{0.95\textheight}
\tablecaption{Selected examples of various formulae utilizing the nominal
constants.\label{formu}}
\tablehead{
\colhead{Quantity} & \colhead{Units} & \colhead{Conversion expression} & \colhead{Numerical relation}
}
\startdata
\multicolumn{4}{c}{Kepler's third law for two-body problem: binaries}\\
\noalign{\smallskip}\tableline\noalign{\smallskip}
& (m) &($\GMnom/(4\pi^2))^{1/3}(86400)^{2/3}$ &$2.927699\cdots \times 10^{9}\ P^{2/3} \left( M_1 + M_2 \right)^{1/3}$\\
semi-major axis $a$ & (au) &($\GMnom/(4\pi^2))^{1/3}(86400)^{2/3}/au$& $0.01957046\cdots\ P^{2/3} \left( M_1 + M_2 \right)^{1/3}$\\
 & ($\Rnom$) &($\GMnom/(4\pi^2))^{1/3}(86400^{2/3}/\Rnom)$&$4.208278\cdots\ P^{2/3} \left(M_1 + M_2 \right)^{1/3}$\\
\noalign{\smallskip}\tableline\noalign{\smallskip}
\multicolumn{4}{c}{Double-lined spectroscopic binaries}\\
\noalign{\smallskip}\tableline\noalign{\smallskip}
stellar masses $M_{1,2}$&\Mnom&
$(86400 \times 1000^3)/(2\pi\GMnom)$&$M_{1,2}\sin^3 i
= 1.036149\cdots \times 10^{-7} K_{2,1}(K_1+K_2)^2 P(1-e^2)^{3/2}$ \\
Projected orbital sizes $a_{1,2}$&\Rnom&
$(86400\times1000)/(2\pi\Rnom)$&
$a_{1,2}\sin i = 0.01976569\cdots K_{1,2}P(1-e^2)^{1/2}$\\
semi-major axis $a=a_1+a_2$  &\Rnom&$(86400\times1000)/(2\pi\Rnom)$&
   $a \sin i = 0.01976569\cdots (K_1+K_2)P(1-e^2)^{1/2}$ \\
\noalign{\smallskip}\tableline\noalign{\smallskip}
\multicolumn{4}{c}{Single-lined spectroscopic binaries}\\
\noalign{\smallskip}\tableline\noalign{\smallskip}
Mass function $f_{1,2}(M_1,M_2)$&\Mnom&$(86400\times1000^3)/(2\pi\GMnom)$
& $f_{1,2}(M_1,M_2) = 1.036149\cdots \times 10^{-7} K_{1,2}^3 P(1-e^2)^{3/2}$\\
Mass of invisible component $M_{2,1}$&\Mnom&$1000(86400/(2\pi\GMnom))^{1/3}$
&$M_{2,1}\sin i =0.004696858\cdots\ K_{1,2}(M_1+M_2)^{2/3}P^{1/3}(1-e^2)^{1/2}$\\
\noalign{\smallskip}\tableline\noalign{\smallskip}
\multicolumn{4}{c}{Various formulae\ related to individual stars}\\
\noalign{\smallskip}\tableline\noalign{\smallskip}
log of surface gravity $g$ & log (cm s$^2$) &$\log(10^6\GMnom)-2\log(100\Rnom)$
& $4.438068\cdots + \log M - 2 \log R$ \\
\noalign{\smallskip}\noalign{\smallskip}
Stellar bolometric magnitude $M_{\rm Bol}$&mag&
$2.5\log(L_0/\Lnom)-2.5\log(4\pi\sigma\Rnom^2/\Lnom)$&
$M_{\rm Bol} = 42.3532632(25) - 5\log R - 10\log T_{\rm eff}$\\
\noalign{\smallskip}\noalign{\smallskip}
Linear stellar radius $R$ from angular diameter&\Rnom&
(1pc/\Rnom) $au$ ($\pi$/180)(1/3600)/2
& $R = 107.5161\cdots\ \theta\,p^{-1}$\\
\noalign{\smallskip}\noalign{\smallskip}
Equatorial rotational velocity $V_{\rm eq}$ & km\,s$^{-1}$&$2\pi\Rnom/(1000 \times 86400)$&
$V_{\rm eq} = 50.59273\cdots\ R\,P_{\rm rot}^{-1}$\\
\noalign{\smallskip}\noalign{\smallskip}
Keplerian velocity for given mass and radius $V_{\rm Kepler}$& km\,s$^{-1}$&
0.001(\GMnom/\Rnom)$^{1/2}$&$V_{\rm Kepler} = 436.7620\cdots\ \sqrt{M/R}$ \\
\enddata
\end{deluxetable}

\end{document}